\documentclass[runningheads]{CMSIM}

\usepackage{graphicx} 
\usepackage{amsmath,amssymb}                      

\renewcommand{\thefootnote}

\title*{Are credit ratings time-homogeneous and Markov?}

\titlerunning{Are credit ratings time-homogeneous and Markov?}

\author{%
  Pedro Lencastre\inst{1,2}, 
  Frank Raischel\inst{3}, 
  Pedro G.~Lind\inst{4},
  Tim Rogers \inst{5}
}

\authorrunning{Lencastre et al}
\institute{%
  ISCTE-IUL, Av.~For\c{c}as Armadas, 1649-026 Lisboa, Portugal\\
  (e-mail: {\tt pedro.lencastre.silva@gmail.com})
  \and
  Mathematical Department, FCUL, University of Lisbon, 1749-016 Lisbon, 
  Portugal\\
  \and
  Instituto Dom Luiz, University of Lisbon, 
  1749-016 Lisbon, Portugal\\
  (e-mail: {\tt raischel@cii.fc.ul.pt})
  \and
  ForWind and Institute of Physics, University of Oldenburg,
  Ammerl\"ander Heerstrasse 136, DE-26111 Oldenburg, Germany\\
  (e-mail: {\tt pedro.g.lind@forwind.de})
  \and
  Centre for Networks and Collective Behaviour,
  Department of Mathematical Sciences, University of Bath,
  Claverton Down, BA2 7AY, Bath, UK
}

\begin{document}
\thispagestyle{empty}
\maketitle             
\setlength{\leftskip}{0pt}
\setlength{\headsep}{16pt}

\begin{abstract}
We introduce a simple approach for testing the reliability of 
homogeneous generators and the  Markov property of the
stochastic processes
underlying empirical time series of credit ratings. We analyze open 
access data provided by Moody's  and show that the validity of these 
assumptions - existence of
a homogeneous generator and Markovianity - is not always guaranteed.
Our analysis is based on a comparison between empirical transition 
matrices aggregated over fixed time windows and candidate transition 
matrices generated from measurements taken over shorter periods.
Ratings  are widely used in credit risk, and are a 
key element in risk assessment; our results provide a tool for
quantifying confidence in predictions extrapolated from rating time 
series.
\end{abstract}

\section{Motivation and Scope}

$ \quad \, $After the Basel II accord in 2004 \cite{basel2}, ratings became an increasingly 
important instrument in Credit Risk, as they 
allow banks to base their 
capital requirements on internal as well as external rating systems. 
These ratings became instrumental in evaluating the risk of a bond or loan 
and in the calculation of the Value at Risk. 
As such, it is often desirable to quantify the uncertainty in these ratings, 
and predict the likelihood that an institution will be upgraded or downgraded in 
the near future. A common technique is to aggregate credit rating transition
data over yearly or quarterly periods, and to model future transitions using 
these data. However, to be reliably 
the ratings' evolution must obey particular features which we show below 
can be evaluated through 
analysis of the data published by rating agencies.
Two sufficient properties for accepting the empirical data as a reliable 
indicator of future rating evolution are the existence of one generator
and Markovianity.

The representation of the evolution of a time-continuous process by an 
aggregated transition matrix will not be adequate if the underlying process 
is not Markov.
Moreover, if there is no generator associated to the transition matrix,
the process underlying the ratings is not continuous.
Different techniques to estimate a transition matrix from a 
finite sample of data should be employed depending on whether the process is 
time-homogeneous or not\cite{Charitos2008computing,Lando2002}.
Theoretically, both the Markov and the time-homogeneous assumptions simplify 
considerably the models in question\cite{weissbach2009homogeneoustest}, but 
typically only the latter is at times dropped in order to build a more 
general theoretical framework.

In this paper we test how good both assumptions are in different periods of 
time for a homogeneous rating class in Moody's database. 
We compare transition matrices calculated under different assumptions 
and show that the quality of the time-homogeneous and Markov assumptions 
change considerably in time. Moreover, we argue that the wide fluctuations
of the assumptions' quality may on the one hand provide evidence for detecting
discontinuities in the rating process, e.g.~when establishing new evaluation
criteria for a bank rating, and, on the other hand,  can be taken 
as a tool for ascertaining how complete and trustable such rating criteria are.

We start in Sec.~\ref{sec:empirical_data} by describing the empirical 
data collected from Moody's and in Sec.~\ref{sec:findings} 
we describe how to test the validity of both the homogeneity
and Markovianity assumptions.
Section \ref{sec:conclusions} concludes the paper and presents some
discussion of our results in the light of finance rating procedures.

\section{Data: Six Years of Rating Transitions in Europe}
\label{sec:empirical_data}

$ \quad \, $The data analyzed in this paper is publicly available data that Moody's needs to disclose and keep
publicly available in compliance with Rule 17g-2(d)(3) of US. SEC
regulations \cite{Database}. 

The rating time series of each bank has a sample frequency of one day,
starting in January 1st 2007 and ending in January 1st of 2013. The data sample 
is the set of rating histories from the banks, in European 
countries, that had a rating at the final date. 
Each value indicates the rating class, according to the so-called 
\textit{Banking Financial Strength}\cite{moody2004definitions}, at which 
the bank is evaluated at that particular day. 
\begin{figure}[htb]
\centering
\includegraphics[width=0.95\textwidth]{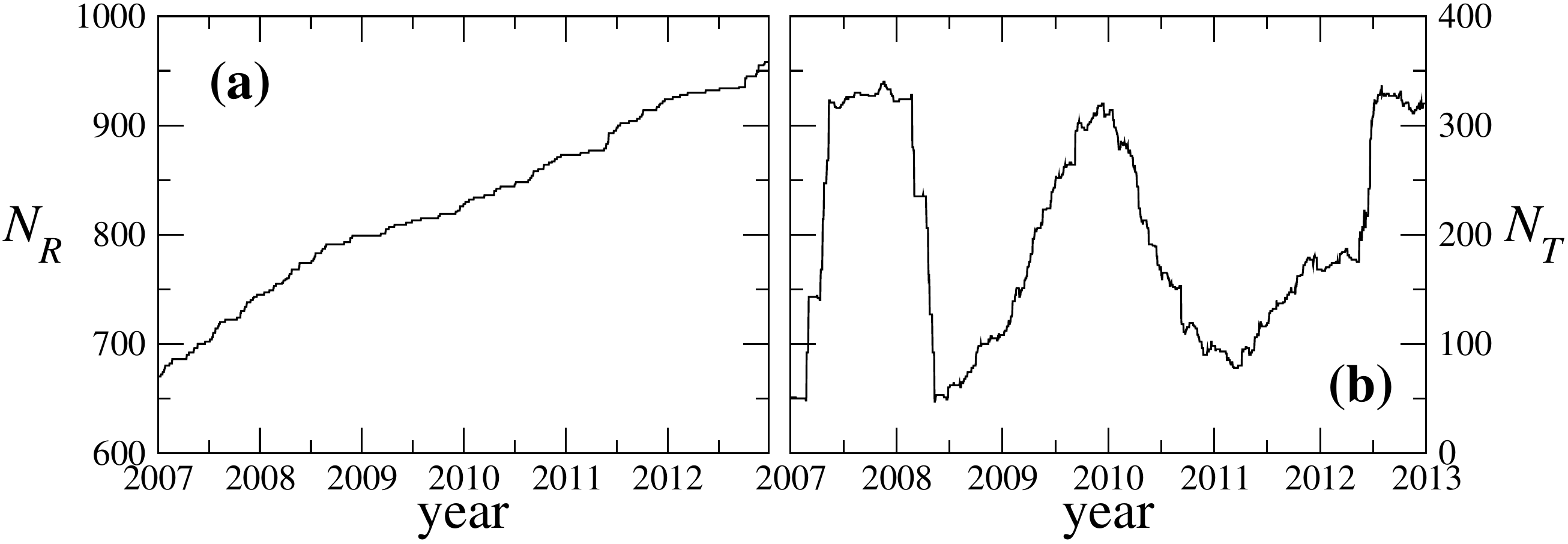}
\caption{\protect
         {\bf (a)} Number of bank entities in Moody's data sample as a 
                   function of time and 
         {\bf (b)} the number of transitions per bank, computed as 
                   moving averages during one-year periods.}
\label{fig1}
\end{figure}
\begin{figure}[htb]
\centering
\includegraphics[width=0.48\textwidth]{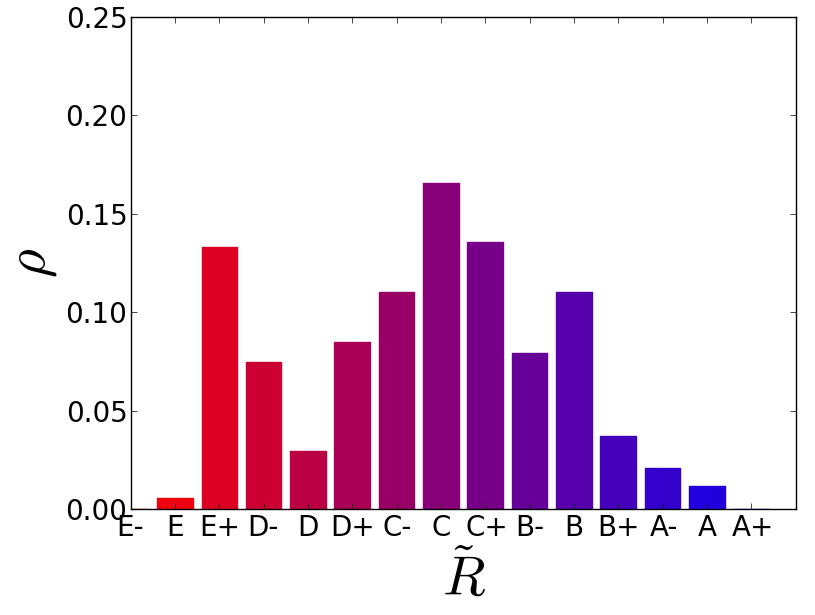}%
\includegraphics[width=0.48\textwidth]{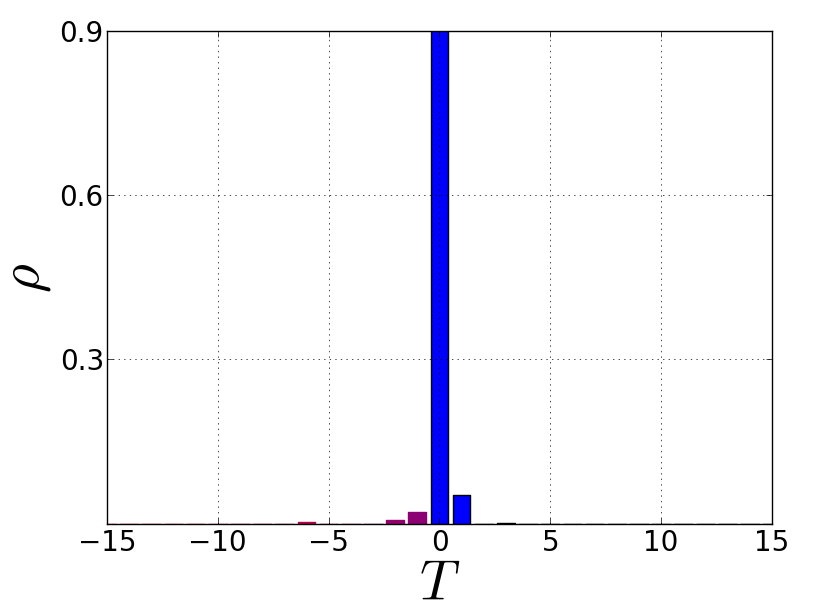}
\includegraphics[width=0.48\textwidth]{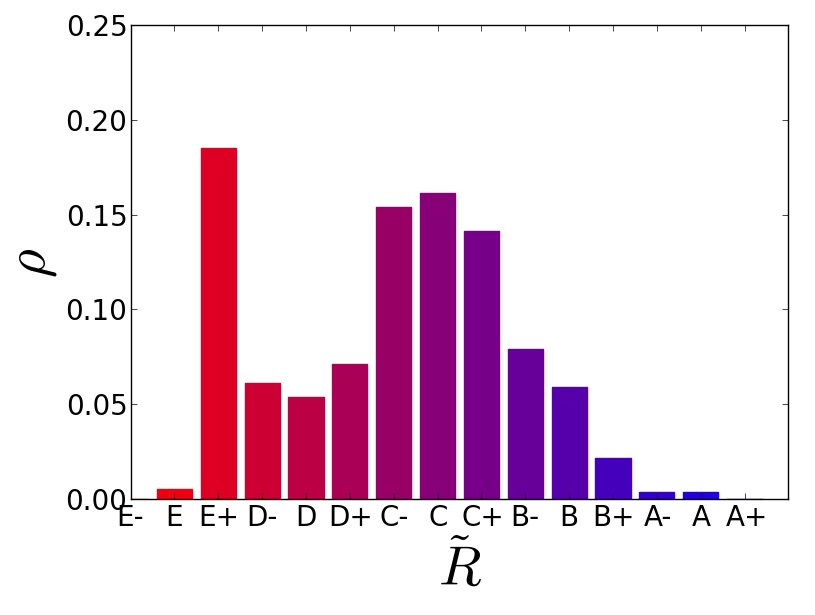}%
\includegraphics[width=0.48\textwidth]{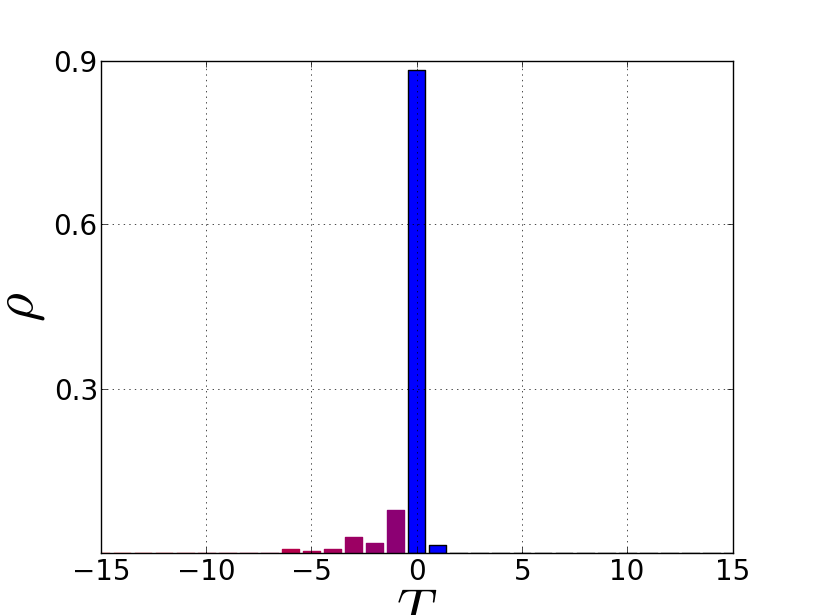}
\includegraphics[width=0.48\textwidth]{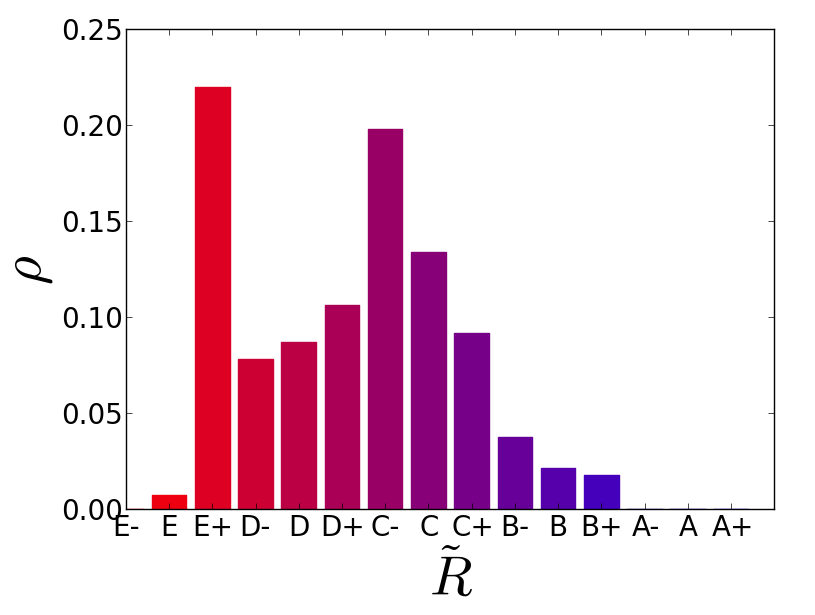}%
\includegraphics[width=0.48\textwidth]{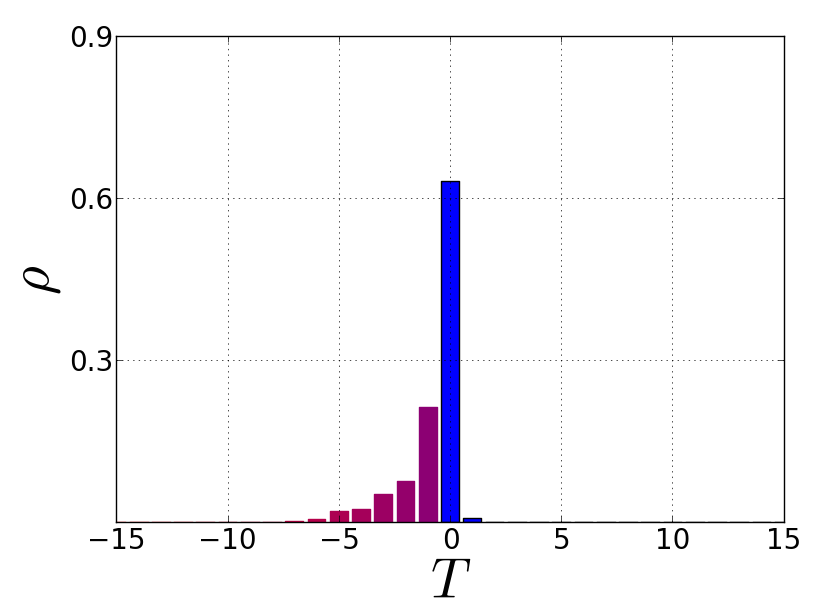}
\caption{\protect
         Illustration of rating histograms for the rating state $\tilde{R}$ (left) 
         and the corresponding rating variations $T=\Delta R$ (right),
         where $R$ is an integer enconding the rating state, ranging
         from $0$ ($E-$) to $14$ ($A+$). Three different
         days are selected: first day of 2007 (first row), 2009(second row)
         and 2010 (third row); cf.~Fig.~\ref{fig3}.}
\label{fig2}
\end{figure}

One first important feature of this rating database is its non-stationary
character, as can be seen in Fig.~\ref{fig1}. The number of banks $N_R$ 
included in the data set increased almost monotonically during the total 
time-span analyzed by us (see Fig.~\ref{fig1}a). 
On  January 1st 2007 there are $N_R=658$ rated companies in the data set, 
and this number increases until 2013 when one registers $N_R=924$ rated banks. 
Therefore, we will consider our measures normalized to the number of banks 
in the database. 

We count in Moody's database a total of 
$N_T=932$ rating transitions, that distribute heterogeneously in time. 
Indeed, the number of 
transitions $N_T$ per bank also changes significantly, with three events
of peaked activity, namely during the year of 2007, at the beginning of
2010 and in the last half year of 2012 (see Fig.~\ref{fig1}b). This will be of importance when
analyzing the evolution of the generator homogeneity and Markovianity
of the corresponding transition matrices.

The rating category is a measure of the capacity of the institution to 
meet its financial obligations and avoids default or government bailout. 
We have $n_s=15$ rating states, 
denoted by the letters $A$ to $E$ in alphabetic order and with the two 
possible extra suffixes, namely $+$ and $-$. 
State $A+$ represents the state corresponding to the best financial health
and less credit risk, followed by $A$, $A-$, $B+$ and so on, until the bottom 
of the scale, $E-$, the state that represents the highest risk level. 
Figure \ref{fig2} shows three plots (left) illustrating the histogram of
rating states at three different time, namely the first day of 2007, 2009
and 2010.

Henceforth, we define $\tilde{R}_i(t)$ as the rating of the bank number $i$ at the 
moment $t$, and 
we map the rating states to an increasing ordered number series: state $\tilde{R}=E+$ 
corresponding to label $R=0$, and state $\tilde{R}=A+$ to label $R=14$. 
With such a labelling it is possible to compute rating increments as
\begin{equation}
T_i(t,\tau) = R_i(t) - R_i(t-\tau) .
\label{transition_process}
\end{equation}
When $T_i(t)>0$ (resp.~$<0$) it means that bank $i$ saw its rating increased 
(resp.~decreased) during the last $\tau$ period of time.
Unless stated otherwise we will use always $\tau = 365$ days.
The plots in the right column of Fig.~\ref{fig2} show the histograms of the
corresponding rating increments at the same three days.

\begin{figure}[htb]
\centering
\includegraphics[width=0.95\textwidth]{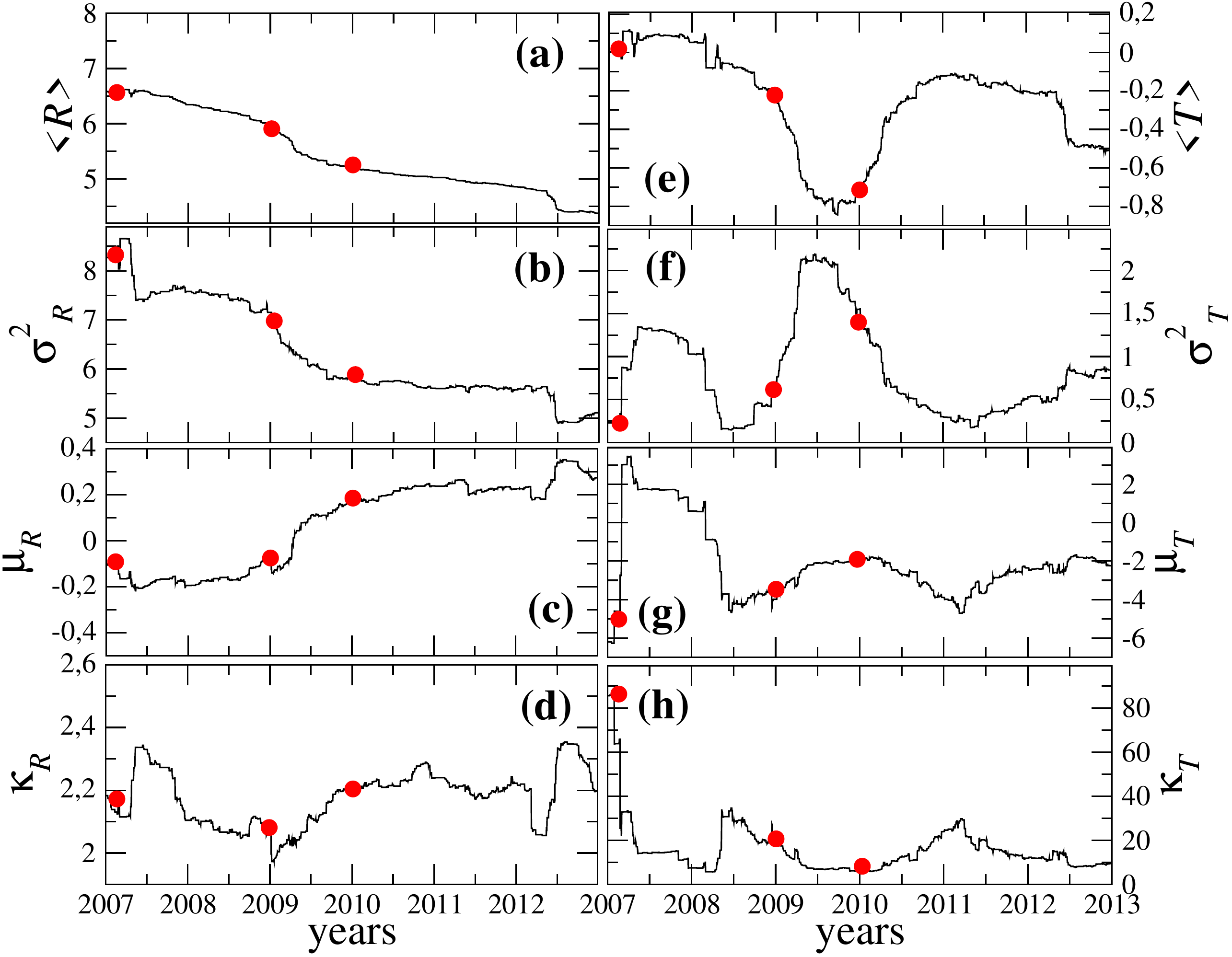}
\caption{\protect
         Evolution of the first statistical moments of 
         {\bf (a-d)} the rating state $R$ distribution and
         {\bf (e-h)} its one-year-increment $T$ distribution.
         From top to bottom: averages $\langle R\rangle$ 
         and $\langle T\rangle$, variances $\sigma^2_R$ and 
         $\sigma^2_T$, skewnesses $\mu_R$ and $\mu_T$, and
         kurtosis $\kappa_R$ and $\kappa_T$. Bullets
         indicate the days when the histograms in 
         Fig.~\ref{fig2} were taken.}
\label{fig3}
\end{figure}

We call henceforth $R(t)$ and $T(t)$  the aggregated processes of the
ratings and rating increments respectively, over all $N_R$ companies 
observed at time $t$.
Figure \ref{fig3} shows the evolution of the first four moments for both
rating distributions (left) and transition distributions (right), with
$\tau = 365$ days.

The average rating $\langle R\rangle$ (Fig.~\ref{fig3}a) has
decreased during most of the six year period records.
We should note however that this is due to the new entries in the database
whose initial rating is typically low, since $\langle T\rangle$ has 
positive periods during the first five years of the recorded set.  

As for the rating variance $\sigma_R$ (Fig.~\ref{fig3}e), after a slight 
increase, it also decreased since the middle of 2007, due to the concentration
of rating states to the lower rating classes ($\langle T \rangle<0$). 
The transitions however exhibit two periods of increased variance $\sigma_T$ 
(Fig.~\ref{fig3}f), which reflect probably the respective increase in the  number 
of transitions (compare with Fig.~\ref{fig1}b).

As the lowest states get more and more dominant, the rating skewness 
$\mu_R$ (Fig.~\ref{fig3}c) increases steadily, until it changes sign around
2008, when transitions become negative on average. These two observations
are consistent with each other: the negative skewness indicates the
large majority of banks being below the average rating which corresponds
to an average decrease of the rating $\langle T\rangle<0$.
It also indicates that there are a few banks highly rated. This observation
together with the observations regarding temporal homogeneity  in the next section will justify
some comments about the objectiveness of rating criteria.

The rating distribution is also typically platykurtic (see Fig.~\ref{fig3}d), 
as its kurtosis is always below three (Gaussian kurtosis), indicating a 
more pronounced flatness around the average of rating distributions.
Concerning the third and fourth moments of transition distributions,
Figs.~\ref{fig3}g and \ref{fig3}h respectively, we 
see large fluctuations during the periods with fewer transitions. 
One can clearly sees a very high kurtosis, and changes in the sign of the 
mean and skewness. 

\section{What is the  Underlying Continuous Process?}
\label{sec:findings}

$ \quad \, $In the following we assume that the set of
rating transitions has  a continuous processes underlying it, 
an assumption which has been the subject of previous investigations without a clear result, see e.g. Ref.~\cite{Israel2001}. Even in  case that there is a continuous process, the corresponding generator may be constant (homogeneous generator) or vary in time (non-homogeneous).

The non-homogeneity is important in the finance context since it limits the range of models that can be used. In particular, it has been argued~\cite{Charitos2008computing} that if we consider time-homogeneity a method for estimating a transition matrix better than the one for the more general case.
The main advantages of this method are to capture very small transition probabilities between two states, even when no transitions occurred between those two states, and to distinguish between transitions within the studied time-frame.
The time-homogeneity condition is also  important to check if the rating philosophies\cite{kiff2013rating,Varsany2007} allegedely used are being correctly followed or not, and they do not hold if criteria by which ratings are ascribed to banks are not constant in time, but vary according to artificial or externally imposed factors\cite{mathis2009}.

Furthermore, another important feature of continuous transition processes 
is their Markovianity.  The Markov property is important if  the current rating of a bank is to be considered a complete indicator of its future  risk. 
In this section we will address both these conditions separately.

\subsection{Testing Time-Homogeneity}
\label{subsec:homogeneity}

$ \quad \, $Mathematically, if a time-continuous Markov process is time-homogeneous 
then there is a constant matrix $\mathbf{Q}$, 
called a generator, solution of
\begin{equation}
\frac{d\mathbf{M}(t)}{dt}=\mathbf{Q}\mathbf{M}(t) ,
\label{MC1}
\end{equation}
where $\mathbf{M}$ is the transition matrix, with entries
$M_{ij}$ given the probability for observing a transition from
state $i$ to state $j$ ($i,j=1,\dots,n_s$).
In other words, a time-continuous process is time-homogeneous
if, being Markov, its transition matrix can be expressed 
as $\mathbf{M}(t) = e^{\mathbf{Q}t}$, and therefore it has a well-defined 
logarithm. 
We take the analogue from ordinary differential equations and 
loosely call $\mathbf{Q}$ the logarithm of $\mathbf{M}$.

\begin{figure}[t]
\centering
\includegraphics[width=0.95\textwidth]{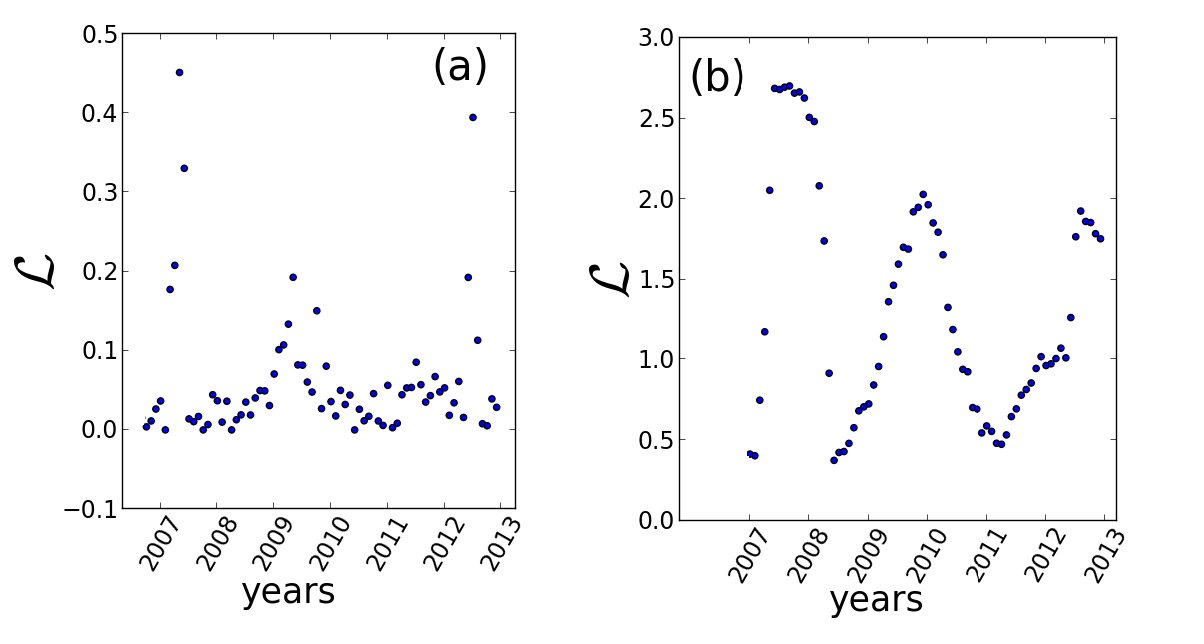}
\caption{\protect 
         Testing for temporal homogeneity: difference between the log-likelihood 
          $\mathcal{L}$ of the 
         transition matrix $\mathbf{M}^{(e)}$ and the transition matrix 
         $\mathbf{M}$ calculated assuming time-homogeneity.
         Both matrices are calculated over a time interval 
         {\bf (a)} one month and 
         {\bf (b)} one year. 
         The log-likelihood was calculated 
         using Eq.~\eqref{eq:log-likelyhood} at the first 
         day of each month from January 2007 to December 2012.}
\label{fig4}
\end{figure} 

The mathematical conditions for the existence of a homogeneous 
generator give a bivalent result\cite{Israel2001,Davies2010} that 
does not take into consideration neither noise generated from finite 
samples
nor how distant an empirical process is from being time-continuous.
Therefore, we neglect several mathematical results that determine if 
a generator exists or not, and assume that the process is Markov and time-continuous.
Being Markov and time-continuous means that there is a generator satisfying 
Eq.~\eqref{MC1} and that it either is constant 
or varies in time.

Next, we estimate the closest constant generator $\mathbf{Q}$ directly from the 
empirical data, compute the associated matrix $\mathbf{M}= 
e^{\mathbf{Q}t}$, and compare it with the empirical transition matrix 
$\mathbf{M}^{(e)}$.
For estimating the generator matrix $\mathbf{Q}$ we follow the approach 
described in Ref.~\cite{Lando2002}, calculating its off-diagonal 
elements as
\begin{equation}
Q_{ij} = \frac{N_T^{(ij)}}{ \int_{t_0}^{t_f} N_R^{(i)}(t) dt} ,
\label{empirical_homogeneous_generator}
\end{equation}
where $N_T^{(ij)}$ represents the number of transitions from $i$ to $j$ 
between the times $t_0$ and $t_f$, and $N_R^{(i)}(t)$ stands for the 
number of banks in  state $i$ at time $t$. 
The diagonal elements $Q_{ii}$ follow from the condition $\sum_j Q_{ij} = 0$.

To compute the distance between a time-homogeneous process and the empirical 
process we compare $\mathbf{M}$ with $\mathbf{M}^{(e)}$, and plot the
statistic:
\begin{equation} \label{eq:log-likelyhood}
\mathcal{L}=
\frac { \sum_{i,j}  N_T^{(ij)} \left(  \log{M_{ij}} - \log{M^{(e)}_{ij}}\right) }{ \sum_{i,j} N_T^{(ij)} } \,.
\end{equation}

This is a log-likelihood ratio; loosely speaking it quantifies  the error 
introduced by making the assumption of time homogeneity. 
The results are shown in Fig.~\ref{fig4}: in panel (a) we aggregate the
data in periods of one month while in panel (b) the aggregation
period is one year.

It can be seen that there are three periods when the time-homogeneity 
condition becomes an insufficient approximation to the dynamics of the process
marked by significant increases in $\mathcal{L}$. 
The first period starts in the early 2007, the second period around the 
middle of 2009, and the third period in the last half of 2012. 
The profile of the time-inhomogeneity is different for each time-period. 
It shows a sharp peak in 2007, concentrated in just a few months, and 
wider  in the other periods. 

These three periods can be better analysed taking also observations from
Fig.~\ref{fig3}.
In 2007 there was an unusually high number of rating transitions, even 
considering that only about $700$ companies were rated at the time. 
In Fig.~\ref{fig3} it can be seen that in this period the variance $\sigma_R$
of the ratings decreased, the skewness $\mu_R$ had slight negative burst, 
and there was an increase in the kurtosis $\kappa_R$. 
As for the statistics of transitions, one can see in this period the 
average $\langle T\rangle$ becoming positive, the skewness ($\mu_T$) 
changing signal and becoming 
positive and the kurtosis ($\kappa_T$) decreasing. The variance 
of $T$ increases, but again that can be explained by the high number of 
rating transitions in that period. 

In late 2009 and early 2010 we have a very different profile. In this period 
the downgrades are the rule, as one can see by the negative values of
$\langle T\rangle$. The relatively low values of $\kappa_T$  and the absolute value of 
$\mu_T$ tells us that this was a general trend, and not a very drastic 
movement by just a few banks. 

In 2012 the scenario is similar to 2010. 
Again there are more downgrades, and this a 
general trend. The companies are now much more clustered, i.e.~with short
dispersion in their ratings, as one can see 
by the low values in $\sigma_R$. 
\begin{figure}[t]
\centering
\includegraphics[width=0.95\textwidth]{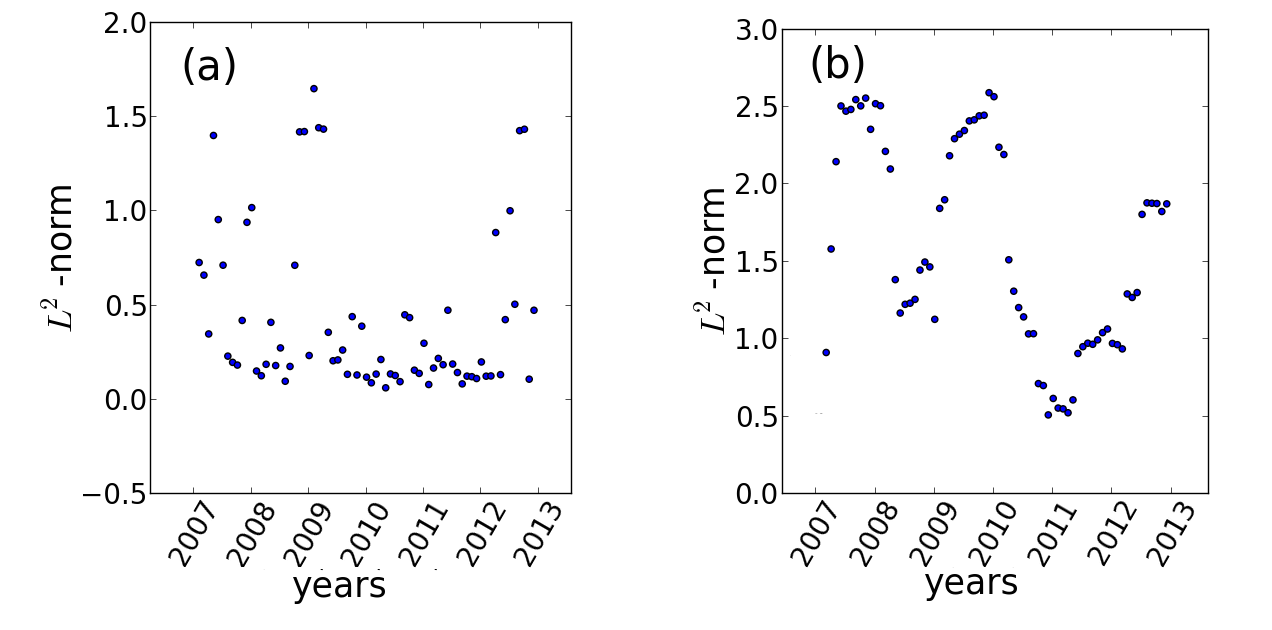}
\caption{\protect 
         Testing Markovianity:
         difference between the empirical
         transition matrix $\mathbf{M}^{(e)}_{0\tau}$ calculated over a time-interval 
         $[0,\tau]$ and the product of the half-period matrices, 
         $\mathbf{M}^{(e)}_{0\tfrac{\tau}{2}}$ and $\mathbf{M}^{(e)}_{\tfrac{\tau}{2}\tau}$, 
         using the $L^2$-norm defined in Eqs.~(\ref{eq:l2}) and (\ref{eq:A}).
         Both matrices are calculated over a time interval of
         {\bf (a)} one month and 
         {\bf (b)} one year. 
         The difference was calculated at the first 
         day of each month between January 2007 and December 2012.}
\label{fig5}
\end{figure}

\subsection{Testing the Markov Hypothesis} 
\label{subsec:markov}

$ \quad \, $Mathematically, a Markov process $x_t$ obeys the following condition:
\begin{equation}
\Pr(x_{t_1} \vert x_{t_2}, x_{t_3}, \dots) = \Pr(x_{t_1} \vert x_{t_2})
\label{mark_condition}
\end{equation}
with $t_1 > t_2  > t_3 > \dots $.
The conditional probability in the right hand-side of 
Eq.~\eqref{mark_condition}, $\Pr(x_{t_1} \vert x_{t_2})$, is exactly specified by  the
transition matrix $\mathbf{M}$. 

The rating process must be assumed to be Markov, otherwise a rating would 
not represent a uniform risk class, as its elements could be distinguished 
according to their previous series of rating states.

From the definition of a Markov process in Eq.~\eqref{mark_condition} it is
straightforward to show that a Markov process also obeys
\begin{equation}
\mathbf{M}_{t_{0}t_{f}} = \prod_{n=1}^{N}\mathbf{M}_{t_{n-1}t_n } ,
\label{Chapman-Kolm}
\end{equation}
where $N$ is the number of subintervals in $[t_0,t_f]$ and
labels $t_it_j$ denote  the time interval $[t_i,t_j]$
considered when determining $\mathbf{M}_{t_it_j}$.
Here we fix $N=2$ and consider two equally spaced intervals with
$\tau\equiv t_f-t_0=1$ month and $\tau=1$ year.
Equation \eqref{Chapman-Kolm} is known as the Chapman-Kolmogorov
equation\cite{risken} and it does not hold in general either when
the process is non-Markov or when we have an insufficiently short  
sample of data.

We will use the Chapman-Kolmogorov equation  as a test indicating 
whe\-ther the rating database of Moody's is Markov.
To that end, we consider empirical matrices $\mathbf{M}^{(e)}_{0\tau}$
computed for one month and one year intervals, and compare it with the 
associated product of the two corresponding half-periods,
$\overline{\mathbf{M}^{(e)}}_{0\tau}=
\mathbf{M}^{(e)}_{0\tfrac{\tau}{2}}\mathbf{M}^{(e)}_{\tfrac{\tau}{2}\tau}$.
For the comparison we now use the $L_2$-norm instead of the 
the $\mathcal{L}$ log-likelihood, since the latter creates 
singularities when dealing with zero entries in the matrices,
and which occur now more frequently.
The $L_2$-norm of the transition matrix is the maximum
singular  value  of $\mathbf{A}$, 
\begin{equation}\label{eq:l2}
\|\mathbf{A}\| = \sigma_{\max} (\mathbf{A}) ,
\end{equation}
and we compute it for as the difference 
\begin{equation}
  \label{eq:A}
\mathbf{A}= \mathbf{M}^{(e)}_{0\tau} - \overline{\mathbf{M}^{(e)}}_{0\tau} \,,  
\end{equation}
where $\|\cdot\|$ represents the usual Euclidian norm.

Results are shown in Fig.~\ref{fig5}.
Clearly, there are two periods when the Markov assumption seems 
less valid. The first period is in early 2007, and the second in the middle 
of 2009, followed by another, less significant increase at the end of 2012.
As said before, this coincides with an abrupt change in the 
statistics of $T$ and $R$.

\section{Discussion and conclusions}
\label{sec:conclusions}

$ \quad \, $We  have addressed time series of credit ratings publicly available at Moody's
online site and 
studied simple ways to compute the validity of the time-homoge-neous and 
Markovianity assumptions. 
We have shown how the accuracy of these assumptions varies 
with time. Naturally, when the Markov assumption fails, so does the 
time-homogeneous assumption, in particular during 2007 and in the latest half of 2009 
and beginning of 2010. In these periods the statistics of the process 
changed considerably. 
In the end of the year of 2012 the accuracy of the time-homogeneous assumption 
is low but the Markov approximation is within the usual fluctuation range. 
In this period there is  a less abrupt change in the statistics of the process.

One must stress that when the Markov assumption does not hold, the ratings 
are not a complete measure of the risk of a given entity, since further 
information besides the actual rating needs to be specified.
Moreover, our results present evidence that perhaps in 2007 new
rating criteria were introduced, imposing a discontinuity in the
series of ratings, or that new rating transition were correlated with 
previous ones, which could support the claim that rating agencies were an 
active part in the crisis that followed. 

Our approach can be 
improved by introducing for instance
a more sophisticated procedure for extracting the histograms for
the ratings and their increments, namely using the kernel based density,
which is known to converge faster to the real distribution than
the usual binning procedure.
From this first approach to investigate Moody's rating database one can now attack the 
embedding problem for the series of transition matrices, where different
generators estimates can be compared. These and other issues will be
addressed elsewhere.

\section*{Acknowledgments}

$ \quad \, $The authors thank 
Funda\c{c}\~ao para a Ci\^encia e a Tecnologia 
for financial support 
under PEst-OE/FIS/UI0618/2011, PEst-OE/MAT/UI0152/2011, SFRH-/BPD/65427/2009 (FR).
This work is part of a bilateral cooperation DRI/-DAAD/1208/2013 
supported by FCT and Deutscher Akademischer Auslandsdienst (DAAD).
PL thanks Global Association of Risks Professionals (GARP)
for the ``Spring 2014 GARP Research Fellowship''.

\bibliographystyle{unsrt}
\bibliography{MatrixBib}

\begin{thebibliography}{10}

\bibitem{basel2}
Basel~Committee on~Banking~Supervision.
\newblock Basel {II}: International convergence of capital measurement and
  capital standards: a revised framework.
\newblock {\em (available at http://bis.org/publ/bcbs107.htm)}, 2004.

\bibitem{Charitos2008computing}
T.~Charitos, P.R. de~Waal, and L.~C. van~der Gaag.
\newblock Computing short-interval transition matrices of a discrete-time
  markov chain from partially observed data.
\newblock {\em Statistics in medicine}, 27(6):905--921, 2008.

\bibitem{Lando2002}
D.~Lando and T.~M. Sk{\o}deberg.
\newblock Analyzing rating transitions and rating drift with continuous
  observations.
\newblock {\em Journal of Banking \& Finance}, 26(2):423--444, 2002.

\bibitem{weissbach2009homogeneoustest}
R.~Wei{\ss}bach, P.~Tschiersch, and C.~Lawrenz.
\newblock Testing time-homogeneity of rating transitions after origination of
  debt.
\newblock {\em Empirical Economics}, 36(3):575--596, 2009.

\bibitem{Database}
https://www.moodys.com/pages/reg001004.aspx.

\bibitem{moody2004definitions}
Moody's~Investors Service.
\newblock Moody's rating symbols \& definitions.
\newblock {\em Report}, 79004(08):1--52, 2004.

\bibitem{Israel2001}
R.B. Israel, J.S. Rosenthal, and J.Z. Wei.
\newblock Finding generators for markov chains via empirical transition
  matrices, with applications to credit ratings.
\newblock {\em Mathematical Finance}, 11(2):245--265, 2001.

\bibitem{kiff2013rating}
J.~Kiff, M.~Kisser, and L.~Schumacher.
\newblock An inspection of the through-the-cycle rating methodology.
\newblock {\em IMF Working Paper}, 2013.

\bibitem{Varsany2007}
Z.~Varsanyi.
\newblock Rating philosophies: some clarifications.
\newblock {\em Report}, 2007.

\bibitem{mathis2009}
J.~Mathis, J.~McAndrews, and J.-C. Rochet.
\newblock Rating the raters: are reputation concerns powerful enough to
  discipline rating agencies?
\newblock {\em Journal of Monetary Economics}, 56(5):657--674, 2009.

\bibitem{Davies2010}
EB~Davies.
\newblock Embeddable markov matrices.
\newblock {\em Electronic Journal of Probability}, 15:1474--1486, 2010.

\bibitem{risken}
H.~Risken.
\newblock {\em The Fokker-Planck Equation}.
\newblock Springer, Berlin, 2nd edition, 1989.

\end{thebibliography}

\end{document}